\documentclass[11pt]{elsart}
\usepackage{amsfonts}
\usepackage{slashbox}
\usepackage{graphicx}
\usepackage{amssymb,amsmath}
\usepackage{mathrsfs} 

\begin{document}

\begin{frontmatter}
\title{Subgame perfect implementation: \\A new result}
\author{Haoyang Wu\corauthref{cor}}
\corauth[cor]{Wan-Dou-Miao Research Lab, Shanghai, China.}
\ead{hywch@mail.xjtu.edu.cn} \ead{Tel: 86-18621753457}

\begin{abstract} This paper concerns what will happen
if quantum mechanics is concerned in subgame perfect implementation.
The main result is: When additional conditions are satisfied, the
traditional characterization on subgame perfect implementation shall
be amended by virtue of a quantum stage mechanism. Furthermore, by
using an algorithmic stage mechanism, this amendment holds in the
macro world too.
\end{abstract}
\begin{keyword}
Mechanism design; Subgame perfect implementation; Quantum game
theory.
\end{keyword}
\end{frontmatter}

\section{Introduction}
The theory of mechanism design plays an important role in economics.
Maskin \cite{Maskin1999} provides an almost complete
characterization of social choice rules that are Nash implementable
when the number of agents is at least three. Moore and Repullo
\cite{MR1988}, Abreu and Sen \cite{AS1990} gave the framework of
subgame perfect implementation. Vartiainen \cite{Vartiainen2007}
proposed a full characterization on subgame perfect implementation.

Recently, Wu \cite{qmd2011} propose that when an additional
condition is satisfied, the sufficient conditions for Nash
implementation shall be amended by virtue of a quantum mechanism.
Although current experimental technologies restrict the quantum
mechanism to be commercially available \cite{Ladd2010}, Wu
\cite{sim2011} propose an algorithmic mechanism by which agents can
benefit from the quantum mechanism immediately when the number of
agents is not very large (e.g., less than 20). Furthermore, the
traditional results on two-agent Nash implementation and Bayesian
implementation are revised \cite{two2011, Bayes2011}.

Following the aforementioned works, this paper investigates what
will happen if the quantum mechanics is generalized to subgame
perfect implementation. The rest of this paper is organized as
follows: Section 2 recalls preliminaries of subgame perfect
implementation given by Abreu and Sen \cite{AS1990}. In Section 3, a
novel condition \emph{multi}-$\alpha$ is defined. In addition, we
give an example of a Pareto-inefficient social choice correspondence
(SCC) that can be implemented in subgame perfect equilibrium.
Section 4 and 5 are the main parts of this paper, in which we will
propose quantum and algorithmic stage mechanisms respectively.
Section 6 draws the conclusions.

\section{Preliminaries}
Let $J=\{1,\cdots,N\}$ be the set of individuals. Let $A$ denote the
set of outcomes. The set of preference profiles is denoted by
$\Theta$. For a profile $\theta\in\Theta$, agent $j\in J$ has
preference ordering $R^{j}(\theta)$ on the set $A$. Let
$P^{j}(\theta)$ and $I^{j}(\theta)$ be the strict preference
relation and the indifference relation associated with
$R^{j}(\theta)$, respectively. For each profile $\theta\in\Theta$,
an SCC $f$ picks a non-empty subset $f(\theta)$ of the set $A$. An
extensive form mechanism is an array $\Gamma=(K,P,U,C,h)$. $K$ is
the game tree with origin $n_{0}$. The set of non-terminal nodes of
$K$ is denoted by $M$. The player, information and choice partitions
are denoted $P$, $U$ and $C$ respectively. The function $h$ is the
payoff function and associates with every path in the tree an
element of the set $A$.

For any profile $\theta\in\Theta$, the pair $(\Gamma, \theta)$
constitutes an extensive form game. A pure strategy for player $j$
is a function which specifies a choice at every information set of
player $j$. The set of pure strategies for player $j$ is denoted
$S^{j}$. A pure strategy profile is an $N$-tuple of pure strategies,
one for each player. The set of pure strategy profiles is denoted
$S=S^{1}\times\cdots\times S^{N}$. Let $M^{0}$ denote the set of all
nodes which are roots of subgames in $\Gamma$. A subgame is
identified by its root $m\in M^{0}$. For all $j\in J$ and $s\in S$,
the set $\Sigma(j,s)$ is defined by $\Sigma(j,s)=\{s'\in
S|s'_{i}=s_{i}, i\neq j\}$. Thus, $\Sigma(j,s)$ is the set of
strategy profiles in which players other than $j$ play according to
$s$.

For all nodes $n$ and $s\in S$, let $a(n,s)$ denote the outcome
obtained under $s$, conditional on starting at node $n$. For all
nodes $n$, $s\in S$ and $j\in J$, let
$A(j,n,s)=\{a(n,s')|s'\in\Sigma(j,s)\}$. The strategy profile $s\in
S$ is a \emph{subgame perfect equilibrium} ($SPE$) of the game
$(\Gamma, \theta)$ if for every subgame $m\in M^{0}$ and every agent
$j\in J$, $a(m,s)$ is $\theta$-maximal for player $j$ in $A(j,m,s)$.
Let $SPE(\Gamma, \theta)$ denote the set of subgame perfect
equilibrium \emph{outcomes} of $(\Gamma, \theta)$. Let
$\overline{SPE}(\Gamma, \theta)$ denote the set of subgame perfect
equilibrium \emph{strategy profiles} of $(\Gamma, \theta)$. The SCC
$f$ is subgame perfect implementable if there exists an extensive
form mechanism $\Gamma$ such that $SPE(\Gamma, \theta)=f(\theta)$
for all $\theta\in\Theta$.

An SCC $f$ satisfies \emph{no-veto power} (NVP) with respect to
$B\subseteq A$ if $\forall\theta\in\Theta$, $\forall a\in B$, [$a
R^{j}(\theta)b$ for all $b\in B$ and $j\in K$, where $K\subseteq J$
and $|K|\geq N-1$]$\Rightarrow a\in f(\theta)$. An SCC $f$ satisfies
\emph{monotonicity} if for every pair of profiles $\theta,
\phi\in\Theta$, and for every $a\in f(\theta)$, whenever
$aR^{j}(\theta)b$ implies $aR^{j}(\phi)b$ ($\forall j\in J$), one
has that $a\in f(\phi)$ \cite{Serrano2004}. An SCC $f$ satisfies
Condition $\alpha$ with respect to the set $B\subseteq A$ if range
$f\subseteq B$, and if for all profiles $\theta, \phi\in\Theta$ and
outcomes $a\in f(\theta)-f(\phi)$, there exist a sequence of agents
$j(\theta, \phi; a)\equiv (j(0),\cdots,j(l))$ and a sequence of
outcomes
$a_{0}, a_{1},\cdots,a_{l},a_{l+1}$ in $B$ (where $a_{0}=a$), such that:\\
(i) $a_{k}R^{j(k)}(\theta)a_{k+1}$, $k=0,\cdots,l$;\\
(ii) $a_{l+1}P^{j(l)}(\phi)a_{l}$;\\
(iii) $a_{k}$ is not $\phi$-maximal for $j(k)$ in $B$, $k=0,\cdots,
l$;\\
(iv) if $a_{l+1}$ is $\phi$-maximal in $B$ for all agents except
$j(l)$, then either $l=0$ or $j(l-1)\neq j(l)$.

\textbf{Theorem 1}: If $f$ is subgame perfect implementable, then
$f$ satisfies Condition $\alpha$ with respect to some $B\subseteq
A$. Let $N\geq 3$. If $f$ satisfies Condition $\alpha$ and NVP with
respect to some $B\subseteq A$, then $f$ is subgame perfect
implementable.

To facilitate the following discussion, here we cite the stage
mechanism used to implement $f$ as follows (P298, \cite{AS1990}).\\
\emph{First Stage (Stage 0)}: Each agent $j\in J$ announces a
triplet $(\theta^{j},a^{j},n^{j})$, where $\theta^{j}\in\Theta$,
$a^{j}\in A$,
and $n^{j}$ is a non-negative integer:\\
Rule (1): If $N-1$ agents announce the same pair $\theta, a\in
f(\theta)$, the outcome is $a$, unless the remaining agent $i$
announces $\phi$, where $a\in f(\theta)-f(\phi)$ and $i=j(0)$ in the
sequence
$j(\theta, \phi; a)$. In the latter event, go to Stage 1. \\
Rule (2): In all other cases, the agent who announces the highest
integer (break ties in favor of the agent with the lowest index) can
select
any outcome in $B$. \\
\emph{Subsequent Stages (Stage k, k=1, $\cdots$, l)}: Each agent
$j\in J$ can either raise a ``flag'' or announce a non-negative
integer:\\
Rule (3): If at least $N-1$ agents raise flags, the agent $j(k-1)$
(in the
sequence $j(\theta, \phi; a))$ can select any outcome in $B$.\\
Rule (4): If at least $N-1$ agents announce zero, the outcome is
$a_{k}$, unless $j(k)$ does not announce zero, in which case go to
the next
stage, or, if $k=l$, implement $a_{l+1}$.\\
Rule (5): In all other cases, the agent who announces the highest
integer can select any outcome in $B$.

\section{Condition \emph{multi}-$\alpha$}
\textbf{Definition 1}: An SCC $f$ satisfies Condition
\emph{multi}-$\alpha$ with respect to the set $B\subseteq A$ if
range $f\subseteq B$, and for two profiles $\theta, \phi\in\Theta$,
$a\in f(\theta)-f(\phi)$, there exist $2\leq\Delta\leq N$ sequences
of agents $j^{\delta}(\theta, \phi; a)\equiv
(j^{\delta}(0),\cdots,j^{\delta}(l^{\delta}))$ (where
$1\leq\delta\leq\Delta$, and no agent belongs to two or more
sequences), and $\Delta$ sequences of outcomes
$a^{\delta}_{0},a^{\delta}_{1},\cdots,a^{\delta}_{l^{\delta}},a^{\delta}_{l^{\delta}+1}$
(where $1\leq\delta\leq\Delta$, $a^{\delta}_{0}=a$) in $B$, such
that for every $1\leq\delta\leq\Delta$,\\
(i) $a^{\delta}_{k}R^{j^{\delta}(k)}(\theta)a^{\delta}_{k+1}$,
$k=0,\cdots, l^{\delta}$.\\
(ii)
$a^{\delta}_{l^{\delta}+1}P^{j^{\delta}(l^{\delta})}(\phi)a^{\delta}_{l^{\delta}}$.\\
(iii) $a^{\delta}_{k}$ is not $\phi$-maximal for $j^{\delta}(k)$ in
$B$, $k=0,\cdots,l^{\delta}$.\\
(iv) If $a^{\delta}_{l^{\delta}+1}$ is $\phi$-maximal in $B$ for all
agents except $j^{\delta}(l^{\delta})$, then either $l^{\delta}=0$,
or $j^{\delta}(l^{\delta}-1)\neq j^{\delta}(l^{\delta})$.

\emph{Table 1: A Pareto-inefficient SCC f that satisfies NVP
and Condition $\alpha$.}\\
\begin{tabular}{cccccccc}
 \multicolumn{4}{c}{Profile $\theta_{1}$}&\multicolumn{4}{c}{Profile $\theta_{2}$}\\
 $Alice$&$Apple$&$Lily$ &$Cindy$ &$Alice$&$Apple$&$Lily$ &$Cindy$\\ \hline
 $u_{4}$&$u_{1}$&$u_{4}$&$u_{1}$ &$u_{4}$&$u_{1}$&$u_{3}$&$u_{1}$ \\
 $u_{2}$&$u_{3}$&$u_{3}$&$u_{3}$ &$u_{1}$&$u_{2}$&$u_{1}$&$u_{4}$ \\
 $u_{1}$&$u_{4}$&$u_{1}$&$u_{2}$ &$u_{2}$&$u_{4}$&$u_{2}$&$u_{3}$ \\
 $u_{3}$&$u_{2}$&$u_{2}$&$u_{4}$ &$u_{3}$&$u_{3}$&$u_{4}$&$u_{2}$ \\\hline
 \multicolumn{4}{c}{$f(\theta_{1})=\{u_{1}\}$}&\multicolumn{4}{c}{$f(\theta_{2})=\{u_{2}\}$}\\\hline
\end{tabular}

\textbf{Example 1}: Consider the SCC $f$ specified in Table 1.
\emph{J}=\emph{\{Alice, Apple, Lily, Cindy}\}, $A=\{u_{1}, u_{2},
u_{3}, u_{4}\}$, $\Theta=\{\theta_{1}, \theta_{2}\}$. Let $B=A$. $f$
is Pareto-inefficient from the viewpoints of agents because in
profile $\theta_{2}$, each agent $j\in J$ prefers a Pareto-efficient
outcome $u_{1}$ to $u_{2}\in f(\theta_{2})$. Obviously, $f$
satisfies NVP. Consider $u_{1}\in f(\theta_{1})$, for every $j\in
J$, whenever $u_{1}R^{j}(\theta_{1})b$ implies
$u_{1}R^{j}(\theta_{2})b$ ($b\in A$), but $u_{1}\notin
f(\theta_{2})$. Therefore, $f$ does not satisfy monotonicity and is
not Nash implementable \cite{Maskin1999}.

Now let us check whether $f$ satisfies Condition $\alpha$ and
\emph{multi}-$\alpha$. Consider two profiles $\theta_{1}$ and
$\theta_{2}$, and outcome $u_{1}\in
f(\theta_{1})-f(\theta_{2})$, then $\Delta=2$.\\
1) For the case of $\delta=1$, $l^{1}=1$, $j^{1}(0)=Alice$,
$j^{1}(1)=Apple$; $a^{1}_{0}=u_{1}$, $a^{1}_{1}=u_{3}$,
$a^{1}_{2}=u_{2}$. \\
(i): $a^{1}_{0}R^{j^{1}(0)}(\theta_{1})a^{1}_{1}$,
$a^{1}_{1}R^{j^{1}(1)}(\theta_{1})a^{1}_{2}$, since
$u_{1}R^{Alice}(\theta_{1})u_{3}$, $u_{3}R^{Apple}(\theta_{1})u_{2}$;\\
(ii): $a^{1}_{2}P^{j^{1}(1)}(\theta_{2})a^{1}_{1}$, since
$u_{2}P^{Apple}(\theta_{2})u_{3}$;\\
(iii): $a^{1}_{0}$ and $a^{1}_{1}$ are not $\theta_{2}$-maximal for
$j^{1}(0)$ and $j^{1}(1)$ in $B$ respectively, since $u_{1}$ and
$u_{3}$ are not $\theta_{2}$-maximal for \emph{Alice} and
\emph{Apple}
in $B$ respectively;\\
(iv): $a^{1}_{2}$ (i.e., $u_{2}$) is not $\theta_{2}$-maximal in $B$
for all agents except $j^{1}(1)$ (i.e., \emph{Apple}).\\
Hence, (i)-(iv) are satisfied.\\
2) For the case of $\delta=2$, $l^{2}=1$, $j^{2}(0)=Lily$,
$j^{2}(1)=Cindy$; $a^{2}_{0}=u_{1}$,
$a^{2}_{1}=u_{2}$, $a^{2}_{2}=u_{4}$. \\
(i): $a^{2}_{0}R^{j^{2}(0)}(\theta_{1})a^{2}_{1}$,
$a^{2}_{1}R^{j^{2}(1)}(\theta_{1})a^{2}_{2}$, since
$u_{1}R^{Lily}(\theta_{1})u_{2}$, and $u_{2}R^{Cindy}(\theta_{1})u_{4}$.\\
(ii): $a^{2}_{2}P^{j^{2}(1)}(\theta_{2})a^{2}_{1}$, since
$u_{4}P^{Cindy}(\theta_{2})u_{2}$;\\
(iii): $a^{2}_{0}$ and $a^{2}_{1}$ are not $\theta_{2}$-maximal for
$j^{2}(0)$ and $j^{2}(1)$ in $B$ respectively, since $u_{1}$ and
$u_{2}$ are not $\theta_{2}$-maximal for \emph{Lily} and
\emph{Cindy} in $B$ respectively.\\
(iv): $a^{2}_{2}$ (i.e., $u_{4}$) is not $\theta_{2}$-maximal in
$B$ for all agents except $j^{2}(1)$ (i.e., \emph{Cindy}). \\
Hence, (i)-(iv) are satisfied. Therefore, $f$ satisfies Condition
$\alpha$ and \emph{multi}-$\alpha$.

Given profile $\theta_{2}$, since $u_{1}$ is Pareto-efficient for
all agents, it looks reasonable that in Stage 0 each agent $j\in J$
announces $(\theta_{1}, u_{1},
*)$ in order to obtain $u_{1}$ by rule (1) (where $*$ stands for
any legal value). But according to Line 19 and the last line of Page
294 \cite{AS1990}, agent $j^{1}(0)=Alice$ has incentives to
unilaterally deviate from $(\theta_{1}, u_{1}, *)$ to $(\theta_{2},
*, *)$ in order to obtain $u_{4}$, which is
$\theta_{2}$-maximal for $Alice$ in $B$ (See Page 294, the third
line to the last, \cite{AS1990}). Similarly, agent $j^{2}(0)=Lily$
also has incentives to unilaterally deviate from $(\theta_{1},
u_{1},
*)$ to $(\theta_{2}, *, *)$ in order to obtain $u_{3}$.

Note that either $Alice$ or $Lily$ can certainly obtain her expected
outcome only if one of them deviates. Since all agents are rational
and self-interested, nobody is willing to give up and let the others
benefit. Therefore, both $Alice$ and $Lily$ will deviate from
$(\theta_{1}, u_{1}, *)$ to $(\theta_{2}, *, *)$. Hence, the
Pareto-efficient outcome $u_{1}$ cannot be implemented in subgame
perfect equilibrium. It should be noted that in the end, rule (2)
will be triggered and the final outcome is uncertain among $B$.

\section{A quantum stage mechanism}
As we have seen in Example 1, although the SCC $f$ is
Pareto-inefficient from the viewpoints of agents, $f$ is subgame
perfect implementable according to the stage mechanism
\cite{AS1990}. Following Ref. \cite{qmd2011}, we will propose a
$(l+1)$-stage quantum mechanism that let the Pareto-efficient
outcome $u_{1}$ be implemented in subgame perfect equilibrium in
profile $\theta_{2}$. The difference between the quantum and
traditional stage mechanisms is the way by which each agent $j\in J$
submits $(\theta^{j}, a^{j}, n^{j})$ to the designer in Stage 0.

\subsection{Assumptions}
According to Eq (4) in Ref. \cite{Flitney2007}, two-parameter
quantum strategies are drawn from the set:
\begin{equation}
\hat{\omega}(\xi,\eta)\equiv \begin{bmatrix}
  e^{i\eta}\cos(\xi/2) & i\sin(\xi/2)\\
  i\sin(\xi/2) & e^{-i\eta}\cos(\xi/2)
\end{bmatrix},
\end{equation}
$\hat{\Omega}\equiv\{\hat{\omega}(\xi,\eta):\xi\in[0,\pi],\eta\in[0,\pi/2]\}$,
$\hat{J}\equiv\cos(\gamma/2)\hat{I}^{\otimes
\Delta}+i\sin(\gamma/2)\hat{\sigma_{x}}^{\otimes \Delta}$, where
$\gamma$ is an entanglement measure, and $\Delta$ is specified in
Condition \emph{multi}-$\alpha$. $\hat{I}\equiv\hat{\omega}(0,0)$,
$\hat{D}_{\Delta}\equiv\hat{\omega}(\pi,\pi/\Delta)$,
$\hat{C}_{\Delta}\equiv\hat{\omega}(0,\pi/\Delta)$. Denote by
$\mathbb{Z}_{+}$ the set of non-negative integer.

\begin{figure}[!t]
\centering
\includegraphics[height=2.9in,clip,keepaspectratio]{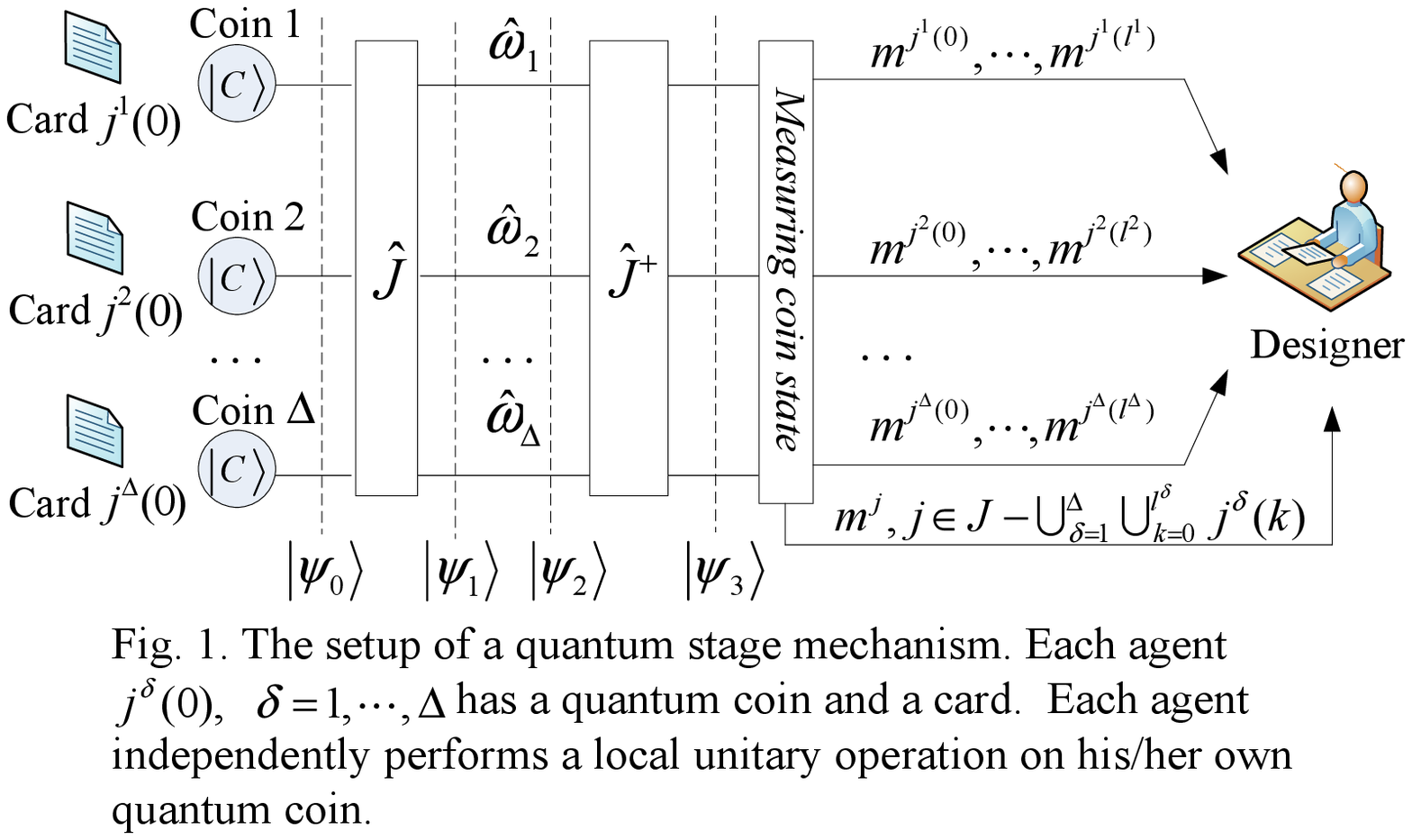}
\end{figure}

Without loss of generality, we assume that:\\
1) Each agent $j^{\delta}(0)$ ($1\leq\delta\leq\Delta$) has a
quantum coin $\delta$ (qubit) and a classical card $j^{\delta}(0)$.
The basis vectors $|C\rangle=(1,0)^{T}$, $|D\rangle=(0,1)^{T}$ of a
quantum coin denote head up and tail
up respectively.\\
2) Each agent $j^{\delta}(0)$ ($1\leq\delta\leq\Delta$)
independently performs a local unitary operation on his/her own
quantum coin. The set of agent $j^{\delta}(0)$'s operation is
$\hat{\Omega}_{\delta}=\hat{\Omega}$. A strategic operation chosen
by agent $j^{\delta}(0)$ is denoted as
$\hat{\omega}_{\delta}\in\hat{\Omega}_{\delta}$. If
$\hat{\omega}_{\delta}=\hat{I}$, then
$\hat{\omega}_{\delta}(|C\rangle)=|C\rangle$,
$\hat{\omega}_{\delta}(|D\rangle)=|D\rangle$; If
$\hat{\omega}_{\delta}=\hat{D}_{\Delta}$, then
$\hat{\omega}_{\delta}(|C\rangle)=|D\rangle$,
$\hat{\omega}_{\delta}(|D\rangle)=|C\rangle$. $\hat{I}$ denotes
``\emph{Not flip}'', $\hat{D}_{N}$ denotes ``\emph{Flip}''.\\
3) The two sides of a card are denoted as Side 0 and Side 1. The
message written on the Side 0 (or Side 1) of card $j^{\delta}(0)$ is
denoted as $card(j^{\delta}(0),0)$ (or $card(j^{\delta}(0),1)$). A
typical card written by agent $j^{\delta}(0)$ is described as
$c_{j^{\delta}(0)}=(card(j^{\delta}(0),0),card(j^{\delta}(0),1))
\in\Theta\times
A\times\mathbb{Z}_{+}\times\Theta\times
A\times\mathbb{Z}_{+}$.\\
4) There is a device that can measure the state of $\Delta$ coins
and send messages to the designer.

\subsection{Condition $\lambda^{SPE}$}
Given $\Delta\geq 2$ agents, consider the payoff to the $\Delta$-th
agent, we denote by $\$_{C\cdots CC}$ the expected payoff when all
$\Delta$ agents choose $\hat{I}$ (the corresponding collapsed state
is $|C\cdots CC\rangle$), and denote by $\$_{C\cdots CD}$ the
expected payoff when the $\Delta$-th agent chooses
$\hat{D}_{\Delta}$ and the first $\Delta-1$ agents choose $\hat{I}$
(the corresponding collapsed state is $|C\cdots CD\rangle$).
$\$_{D\cdots DD}$ and $\$_{D\cdots DC}$ are defined similarly.

\textbf{Definition 2}: Given an SCC $f$ that satisfies
Condition \emph{multi}-$\alpha$, define Condition $\lambda^{SPE}$ as follows: \\
1) $\lambda^{SPE}_{1}$: For the profiles $\theta, \phi$ specified in
Condition \emph{multi}-$\alpha$ and the outcome $a\in
f(\theta)-f(\phi)$, $aR^{j}(\phi)b$ ($b\in f(\phi))$ for every $j\in
J$, and
$aP^{i}(\phi)b$ for at least one $i\in J$.\\
2) $\lambda^{SPE}_{2}$: Consider the payoff to the $\Delta$-th
agent, $\$_{C\cdots CC}>\$_{D\cdots DD}$, i.e., he/she prefers the
expected payoff of a certain outcome (generated by rule (1)) to the
expected payoff of an uncertain outcome (generated by rule (2)). \\
3) $\lambda^{SPE}_{3}$:  Consider the payoff to the $\Delta$-th agent,\\
$\$_{C\cdots CC}>\$_{C\cdots
CD}[1-\sin^{2}\gamma\sin^{2}(\pi/\Delta)]+\$_{D\cdots
DC}\sin^{2}\gamma\sin^{2}(\pi/\Delta)$.

\subsection{Working steps of the quantum stage mechanism}
In the beginning of the traditional stage mechanism, each agent
$j\in J$ directly announces a message $m^{j}$ (i.e., the triplet
$(\theta^{j}, a^{j}, n^{j})$) to the designer, then $N$ agents
participate the stage mechanism as specified by rules (1)-(5). As a
comparison, in the beginning of a quantum stage mechanism, the
message $m^{j}$ ($j\in J$) is generated by another way.

The setup of the quantum stage mechanism is depicted in Fig. 1. The
working steps of a quantum stage mechanism
are described as follows:\\
Step 1: Given an SCC $f$ and a profile $\xi\in\Theta$, if $f$
satisfies Condition \emph{multi}-$\alpha$ and $\lambda^{SPE}$, and
$\xi$ is equal to the profile $\phi$ specified in Condition
$\lambda^{SPE}_{1}$,
then go to Step 3.\\
Step 2: Each agent $j\in J$ sets $m^{j}=(\theta^{j},a^{j},n^{j})$
(where $\theta^{j}\in\Theta$, $a^{j}\in A$,
$n^{j}\in\mathbb{Z}_{+}$). Go to Step 9.\\
Step 3: The state of each quantum coin $\delta$
($1\leq\delta\leq\Delta$) is set as $|C\rangle$. The initial state
of the $\Delta$ quantum coins is
$|\psi_{0}\rangle=\underbrace{|C\cdots CC\rangle}\limits_{\Delta}$.\\
Step 4: Each agent $j^{\delta}(0)$ ($1\leq\delta\leq\Delta$) sets
$c_{j^{\delta}(0)}=((\theta, a,*),(\phi, *, *))$, where $\theta,
\phi, a$ are specified in Condition $\lambda^{SPE}_{1}$. Each agent
$j\in J-\bigcup_{\delta=1}^{\Delta}j^{\delta}(0)$ sets
$c_{j}=((\theta, a, *)$, $(\theta^{j}, a^{j}, n^{j}))$.\\
Step 5: Let $\Delta$ quantum coins be entangled by $\hat{J}$.
$|\psi_{1}\rangle=\hat{J}|\psi_{0}\rangle$.\\
Step 6: Each agent $j^{\delta}(0)$ independently performs a local
unitary operation $\hat{\omega}_{\delta}$ on his/her own quantum
coin.
$|\psi_{2}\rangle=[\hat{\omega}_{1}\otimes\cdots\otimes\hat{\omega}
_{\Delta}]\hat{J}|\psi_{0}\rangle$.\\
Step 7: Let $\Delta$ quantum coins be disentangled by $\hat{J}^{+}$.
$|\psi_{3}\rangle=\hat{J}^{+}[\hat{\omega}_{1}\otimes\cdots\otimes
\hat{\omega}_{\Delta}]\hat{J}|\psi_{0}\rangle$.\\
Step 8: The device measures the state of $\Delta$ quantum coins. For
every $1\leq\delta\leq\Delta$, if the state of quantum coin $\delta$
is $|C\rangle$ (or $|D\rangle$), then the device sends
$card(j^{\delta}(k),0)$ (or $card(j^{\delta}(k),1)$) as
$m^{j^{\delta}(k)}$ ($0\leq k\leq l^{\delta}$) to the designer. For
each agent $j\in
J-\bigcup_{\delta=1}^{\Delta}\bigcup_{k=0}^{l^{\delta}}j^{\delta}(k)$,
the device sends
$m^{j}=(\theta, a, n^{j})$ to the designer.\\
Step 9: The designer receives the overall message
$m=(m^{1},\cdots,m^{N})$.\\
Step 10: The agents and the designer continue to participate the
rules (1)-(5) of the traditional stage mechanism. END.

\subsection{New result for subgame perfect implementation}
\textbf{Proposition 1:} For $N\geq 3$, consider an SCC $f$ that
satisfies NVP and Condition $\alpha$, if $f$ satisfies
\emph{multi}-$\alpha$ and $\lambda^{SPE}$, then $f$ is not
subgame perfect implementable by using the quantum stage mechanism.\\
\textbf{Proof}: Consider the profiles $\theta, \phi$ specified in
$\lambda^{SPE}_{1}$. Since $f$ satisfies Condition
\emph{multi}-$\alpha$ and $\lambda^{SPE}$, then the quantum stage
mechanism enters Step 3, and there exist $2\leq\Delta\leq N$
sequences of agents that satisfy (i)-(iv) of Condition
\emph{multi}-$\alpha$. At first sight, each agent $j^{\delta}(0)$
($1\leq\delta\leq\Delta$) has incentives to unilaterally deviate
from $(\theta, a, *)$ to $(\phi, *, *)$ in order to obtain his/her
$\phi$-maximal outcome (as we have seen in Page 294,
\cite{AS1990}).\\
Consider the payoff to the $\Delta$-th agent (denoted as
\emph{Laura}), when she plays $\hat{\omega}(\xi,\eta)$ while the
first $\Delta-1$ agents play
$\hat{C}_{\Delta}=\hat{\omega}(0,\pi/\Delta)$, according to Refs.
\cite{qmd2011, Flitney2007},
\begin{align*}
  \langle\$_{Laura}\rangle=&\$_{C\cdots CC}\cos^{2}(\xi/2)[1-\sin^{2}\gamma
  \sin^{2}(\eta-\pi/\Delta)]\\
  +&\$_{C\cdots CD}\sin^{2}(\xi/2)[1-\sin^{2}\gamma
  \sin^{2}(\pi/\Delta)]\\
  +&\$_{D\cdots DC}\sin^{2}(\xi/2)\sin^{2}\gamma \sin^{2}(\pi/\Delta)\\
  +&\$_{D\cdots DD}\cos^{2}(\xi/2)\sin^{2}\gamma
  \sin^{2}(\eta-\pi/\Delta)
\end{align*}
Since Condition $\lambda^{SPE}_{2}$ is satisfied, then $\$_{C\cdots
CC}>\$_{D\cdots DD}$, $Laura$ chooses $\eta=\pi/\Delta$ to minimize
$\sin^{2}(\eta-\pi/\Delta)$. As a result,
\begin{align*}
  \langle\$_{Laura}\rangle=&\$_{C\cdots CC}\cos^{2}(\xi/2)\\
  +&\$_{C\cdots CD}\sin^{2}(\xi/2)[1-\sin^{2}\gamma
  \sin^{2}(\pi/\Delta)]\\
  +&\$_{D\cdots DC}\sin^{2}(\xi/2)\sin^{2}\gamma \sin^{2}(\pi/\Delta)
\end{align*}
Since Condition $\lambda^{SPE}_{3}$ is satisfied, then $Laura$
prefers $\xi=0$, which leads to
$\langle\$_{Laura}\rangle=\$_{C\cdots CC}$. In this case,
$\hat{\omega}_{Laura}(\xi,\eta)=\hat{\omega}(0,\pi/\Delta)=\hat{C}_{\Delta}$.\\
By symmetry, in Step 6, each agent $j^{\delta}(0)$
($1\leq\delta\leq\Delta$) chooses
$\hat{\omega}_{\delta}=\hat{C}_{\Delta}$. In Step 8, the collapsed
state of each quantum coin $\delta$ is $|C\cdots CC\rangle$, and
$m^{j}=(\theta, a, *)$ for each agent $j\in J$. No agent
$j^{\delta}(0)$ ($1\leq\delta\leq\Delta$) is willing to deviate from
$(\theta, a, *)$ to $(\phi, *, *)$ in the quantum stage mechanism.
Consequently, in profile $\phi$, $a\in f(\theta)-f(\phi)$ can be
implemented in subgame perfect equilibrium by rule (1). Therefore,
$f$ is not subgame perfect implementable. $\quad\quad\square$

Let us reconsider Example 1. $f$ satisfies Condition
\emph{multi}-$\alpha$ and $\lambda^{SPE}_{1}$. $\theta=\theta_{1}$,
$\phi=\theta_{2}$. The quantum stage mechanism will enter Step 3
when the true profile is $\theta_{2}$. Each agent $j\in\{Alice,
Lily\}$ sets $c_{j}=((\theta_{1},u_{1},*),(\theta_{2},*, *))$. Each
agent $j\in\{Apple, Cindy\}$ sets
$c_{j}=((\theta_{1},u_{1},*),(\theta^{j},a^{j}, n^{j}))$. For any
agent $j\in\{Alice, Lily\}$, let her be the last agent. Consider the
payoff to the fourth agent, suppose $\$_{CCCC}=3$ (the corresponding
outcome is $u_{1}$), $\$_{CCCD}=5$ (the corresponding outcome is
$u_{4}$ if $j=Alice$, and $u_{3}$ if $j=Lily$), $\$_{DDDC}=0$ (the
corresponding outcome is $u_{3}$ if $j=Alice$, and $u_{4}$ if
$j=Lily$), $\$_{DDDD}=1$ (the corresponding outcome is uncertain
among $A$). Hence, Condition $\lambda^{SPE}_{2}$ is satisfied, and
Condition $\lambda^{SPE}_{3}$ becomes:
$3\geq5[1-\sin^{2}\gamma\sin^{2}(\pi/2)]$. If
$\sin^{2}\gamma\geq0.4$, Condition $\lambda^{SPE}_{3}$ is satisfied.
According to Proposition 1, in profile $\theta_{2}$, the
Pareto-efficient outcome $u_{1}$ is generated in subgame perfect
equilibrium by using the quantum stage mechanism.

\section{An algorithmic stage mechanism}
Following Ref. \cite{sim2011}, in this section we will propose an
algorithmic stage mechanism to help agents benefit from the quantum
stage mechanism immediately. In the beginning, we cite the matrix
representations of quantum states from Ref. \cite{sim2011}.

\subsection{Matrix representations of quantum states}
In quantum mechanics, a quantum state can be described as a vector.
For a two-level system, there are two basis vectors: $(1,0)^{T}$ and
$(0,1)^{T}$. In the beginning, we define:
\begin{equation}
|C\rangle=\begin{bmatrix}
  1\\
  0
\end{bmatrix},\quad \hat{I}=\begin{bmatrix}
  1 & 0\\
  0 & 1
\end{bmatrix},\quad \hat{\sigma}_{x}=\begin{bmatrix}
  0 & 1\\
  1 & 0
\end{bmatrix},
|\psi_{0}\rangle=\underbrace{|C\cdots
CC\rangle}\limits_{\Delta}=\begin{bmatrix}
  1\\
  0\\
  \cdots\\
  0
\end{bmatrix}_{2^{\Delta}\times1}
\end{equation}
\begin{align}
&\hat{J}=\cos(\gamma/2)\hat{I}^{\otimes
\Delta}+i\sin(\gamma/2)\hat{\sigma}_{x}^{\otimes \Delta}\\
&\quad=\begin{bmatrix}
  \cos(\gamma/2) &  &  &  &  &  & i\sin(\gamma/2)\\
   & \cdots  & &  & & \cdots  & \\
   &  &  & \cos(\gamma/2) & i\sin(\gamma/2) &  & \\
   &  &  & i\sin(\gamma/2) & \cos(\gamma/2) &  & \\
   & \cdots  & &  &  & \cdots & \\
  i\sin(\gamma/2) &  &  &  &  &  & \cos(\gamma/2)
\end{bmatrix}_{2^{\Delta}\times2^{\Delta}}
\end{align}
For $\gamma=\pi/2$,
\begin{align}
\hat{J}_{\pi/2}=\frac{1}{\sqrt{2}}\begin{bmatrix}
  1 &  &  &  &  &  & i\\
   & \cdots  & &  & & \cdots  & \\
   &  &  & 1 & i &  & \\
   &  &  & i & 1 &  & \\
   & \cdots  & &  &  & \cdots & \\
  i &  &  &  &  &  & 1
\end{bmatrix}_{2^{\Delta}\times 2^{\Delta}},\quad
\hat{J}^{+}_{\pi/2}=\frac{1}{\sqrt{2}}\begin{bmatrix}
  1 &  &  &  &  &  & -i\\
   & \cdots  & &  & & \cdots  & \\
   &  &  & 1 & -i &  & \\
   &  &  & -i & 1 &  & \\
   & \cdots  & &  &  & \cdots & \\
  -i &  &  &  &  &  & 1
\end{bmatrix}_{2^{\Delta}\times2^{\Delta}}
\end{align}

\subsection{An algorithm that simulates the quantum operations and measurements}
Similar to Ref. \cite{sim2011}, in the following we will propose an
algorithm that simulates the quantum operations and measurements in
the quantum stage mechanism. The amendment here is that now the
inputs and outputs are adjusted to the case of subgame perfect
implementation. The factor $\gamma$ is also set as its maximum
$\pi/2$. For $\Delta$ agents, the inputs and outputs of the
algorithm are illustrated in Fig. 2. The \emph{Matlab} program is
given in Fig. 3.

\begin{figure}[!t]
\centering
\includegraphics[height=3in,clip,keepaspectratio]{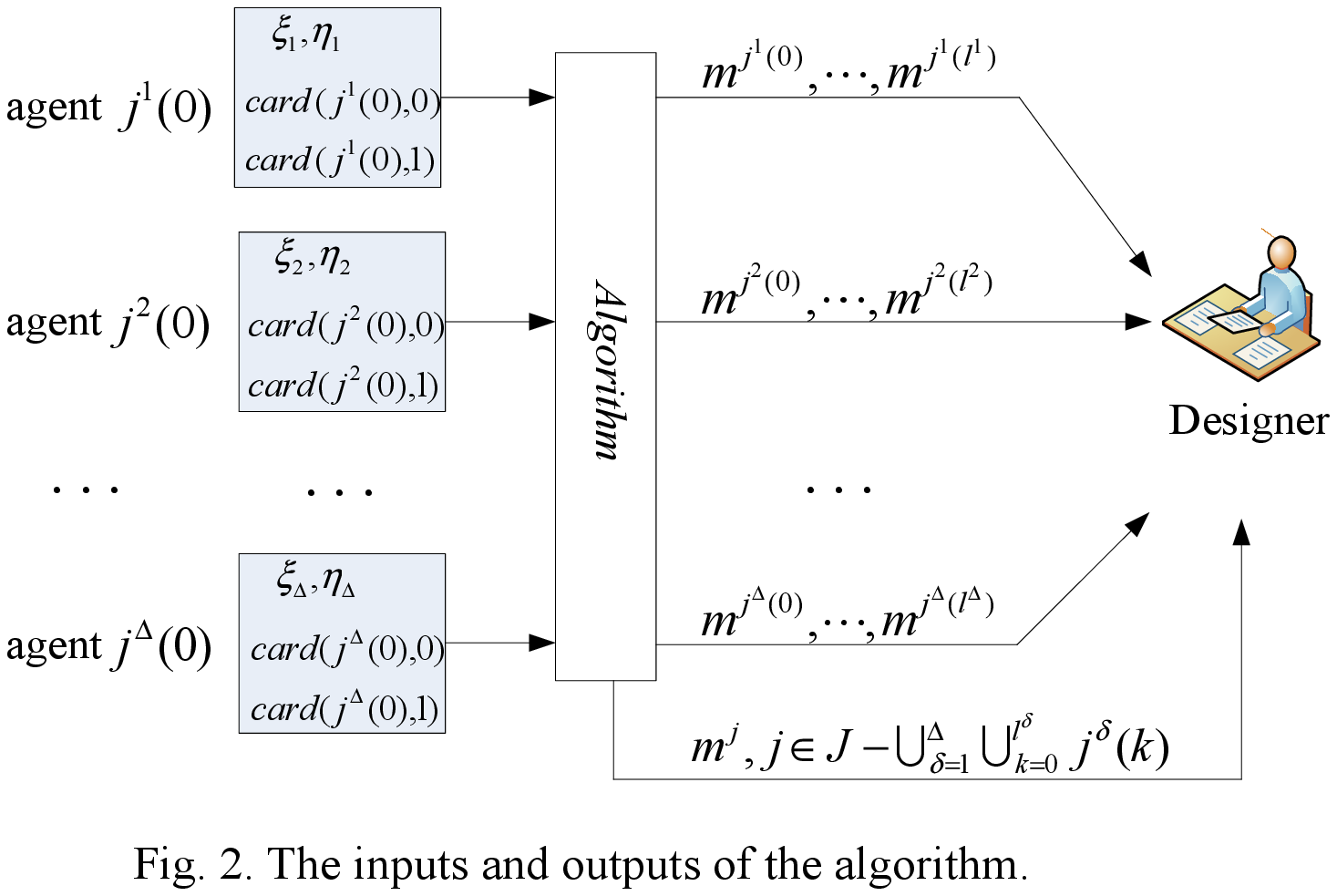}
\end{figure}

\textbf{Inputs}:\\
1) $\xi_{\delta}$, $\eta_{\delta}$, $\delta=1,\cdots,\Delta$: the
parameters of agent $j^{\delta}(0)$'s local operation
$\hat{\omega}_{\delta}$,
$\xi_{\delta}\in[0,\pi],\eta_{\delta}\in[0,\pi/2]$.\\
2) $card(j^{\delta}(0),0), card(j^{\delta}(0),1)$,
$\delta=1,\cdots,\Delta$: the information written on the two sides
of agent $j^{\delta}(0)$'s card, where
$card(j^{\delta}(0),0),card(j^{\delta}(0),1)\in \Theta\times A
\times \mathbb{Z}_{+}$.

\textbf{Outputs}:\\
$m^{j^{\delta}(k)}$ ($\delta=1,\cdots,\Delta$, $k=0,\cdots,
l^{\delta}$): the agent $j^{\delta}(k)$'s message that is sent to
the designer, $m^{j^{\delta}(k)}\in \Theta\times A \times
\mathbb{Z}_{+}$.

\textbf{Procedures of the algorithm}:\\
Step 1: Reading parameters $\xi_{\delta}$ and $ \eta_{\delta}$ from
each agent $j^{\delta}(0)$ ($\delta=1,\cdots, \Delta$) (See Fig. 3(a)).\\
Step 2: Computing the leftmost and rightmost columns of
$\hat{\omega}_{1}\otimes\hat{\omega}_{2}\otimes\cdots\otimes\hat{\omega}_{\Delta}$ (See Fig. 3(b)).\\
Step 3: Computing the vector representation of
$|\psi_{2}\rangle=[\hat{\omega}_{1}\otimes\cdots\otimes\hat{\omega}_{\Delta}]
\hat{J}_{\pi/2}|\psi_{0}\rangle$.\\
Step 4: Computing the vector representation of
$|\psi_{3}\rangle=\hat{J}^{+}_{\pi/2}|\psi_{2}\rangle$.\\
Step 5: Computing the probability distribution
$\langle\psi_{3}|\psi_{3}\rangle$ (See Fig. 3(c)).\\
Step 6: Randomly choosing a ``collapsed'' state from the set of all
$2^{\Delta}$ possible states $\{|C\cdots CC\rangle, \cdots,|D\cdots
DD\rangle\}$ according to the probability
distribution $\langle\psi_{3}|\psi_{3}\rangle$.\\
Step 7: For each agent $j^{\delta}(k)$ ($\delta=1,\cdots,\Delta$,
$k=0,\cdots, l^{\delta}$), the algorithm sends
$card(j^{\delta}(k),0)$ (or $card(j^{\delta}(k),1)$) as the message
$m^{j^{\delta}(k)}$ to the designer if the $\delta$-th basis vector
of the ``collapsed'' state is $|C\rangle$ (or $|D\rangle$).\\
Step 8: For each agent $j\in
J-\bigcup_{\delta=1}^{\Delta}\bigcup_{k=0}^{l^{\delta}}j^{\delta}(k)$,
the algorithm sends $m^{j}=card(j, 0)$ to the designer (See Fig.
3(d)).

\subsection{An algorithmic version of the quantum stage mechanism}
Since the entanglement measure $\gamma$ is set as its maximum
$\pi/2$, the Condition $\lambda^{SPE}$ shall be revised as
$\lambda'^{SPE}$.

\textbf{Definition 6}: Given an SCC $f$ that satisfies
Condition \emph{multi}-$\alpha$, define Condition $\lambda'^{SPE}$ as follows: \\
1) $\lambda'^{SPE}_{1}$ and $\lambda'^{SPE}_{2}$ are the same as
$\lambda^{SPE}_{1}$ and $\lambda^{SPE}_{2}$ respectively.\\
2) $\lambda'^{SPE}_{3}$:  Consider the payoff to the $\Delta$-th
agent, $\$_{C\cdots CC}>\$_{C\cdots
CD}\cos^{2}(\pi/\Delta)]+\$_{D\cdots DC}\sin^{2}(\pi/\Delta)$.

Following Ref. \cite{sim2011}, after quantum operations and
measurements in quantum stage mechanism are replaced by an
algorithm, the quantum stage mechanism shall be revised as an
algorithmic stage mechanism. The working steps of an algorithmic
stage mechanism are described as follows:

Step 1: Given an SCC $f$ and a profile $\xi\in\Theta$, if $f$
satisfies Condition \emph{multi}-$\alpha$ and $\lambda'^{SPE}$, and
$\xi$ is equal to the profile $\phi$ specified in Condition
$\lambda'^{SPE}_{1}$, then go to Step 3.\\
Step 2: Each agent $j\in J$ sets $m^{j}=(\theta^{j},a^{j},n^{j})$
(where $\theta^{j}\in\Theta$, $a^{j}\in A$,
$n^{j}\in\mathbb{Z}_{+}$). Go to Step 6.\\
Step 3: Each agent $j^{\delta}(0)$ ($\delta=1,\cdots,\Delta$) sets
$c_{j^{\delta}(0)}=((\theta, a, *),(\phi, *, *))$, where $\theta,
\phi, a$ are specified in $\lambda'^{SPE}_{1}$; Each agent $j\in
J-\bigcup_{\delta=1}^{\Delta}j^{\delta}(0)$ sets $c_{j}=((\theta, a,
*), (\theta^{j}, a^{j}, n^{j}))$.\\
Step 4: Each agent $j^{\delta}(0)$ ($\delta=1,\cdots,\Delta$)
submits $\xi_{\delta}$, $\eta_{\delta}$, $card(j^{\delta}(0),0)$ and
$card(j^{\delta}(0),1)$ to the algorithm.\\
Step 5: The algorithm runs in a computer and outputs messages
$m^{j}$ ($j\in J$) to the designer.\\
Step 6: The designer receives the overall message
$m=(m^{1},\cdots,m^{N})$.\\
Step 7: The agents and the designer continue to participate the
rules (1)-(5) of the traditional stage mechanism. END.

\textbf{Proposition 2:} For $N\geq 3$, consider an SCC $f$ that
satisfies NVP and Condition $\alpha$, if $f$ satisfies Condition
\emph{multi}-$\alpha$ and $\lambda'^{SPE}$, then $f$ is not
subgame perfect implementable by using the algorithmic stage mechanism.\\

The proof of proposition 2 is straightforward according to
Proposition 1 and Ref. \cite{sim2011}. Note: Although the
algorithmic stage mechanism stems from quantum mechanics, it is
completely \emph{classical} that can be run in a computer. In
addition, Condition $\lambda'^{SPE}$ is also a classical condition.

\section{Conclusions}
This paper follows the series of papers on quantum mechanism
\cite{qmd2011, sim2011, two2011, Bayes2011}. In this paper, the
quantum and algorithmic mechanisms in Refs. \cite{qmd2011, sim2011}
are generalized to subgame perfect implementation. In Example 1, we
show a Pareto-inefficient SCC $f$ that satisfies NVP and Condition
$\alpha$, but does not satisfies monotonicity. According to the
traditional stage mechanism \cite{AS1990}, $f$ can be subgame
perfect implemented. However, by virtue of a quantum stage
mechanism, the Pareto-inefficient SCC $f$ can not be implemented in
subgame perfect equilibrium. Although current experimental
technologies restrict the quantum stage mechanism to be commercially
available, for small-scale cases (e.g., $\Delta\leq 20$
\cite{sim2011}), the algorithmic stage mechanism can help agents
benefit from quantum stage mechanism immediately.

\section*{Acknowledgments}
The author is very grateful to Ms. Fang Chen (\emph{Alice}), Hanyue
Wu (\emph{Apple}), Hanxing Wu (\emph{Lily}) and Hanchen Wu
(\emph{Cindy}) for their great support.

\newpage
\begin{figure}[!t]
\centering
\includegraphics[height=8in,clip,keepaspectratio]{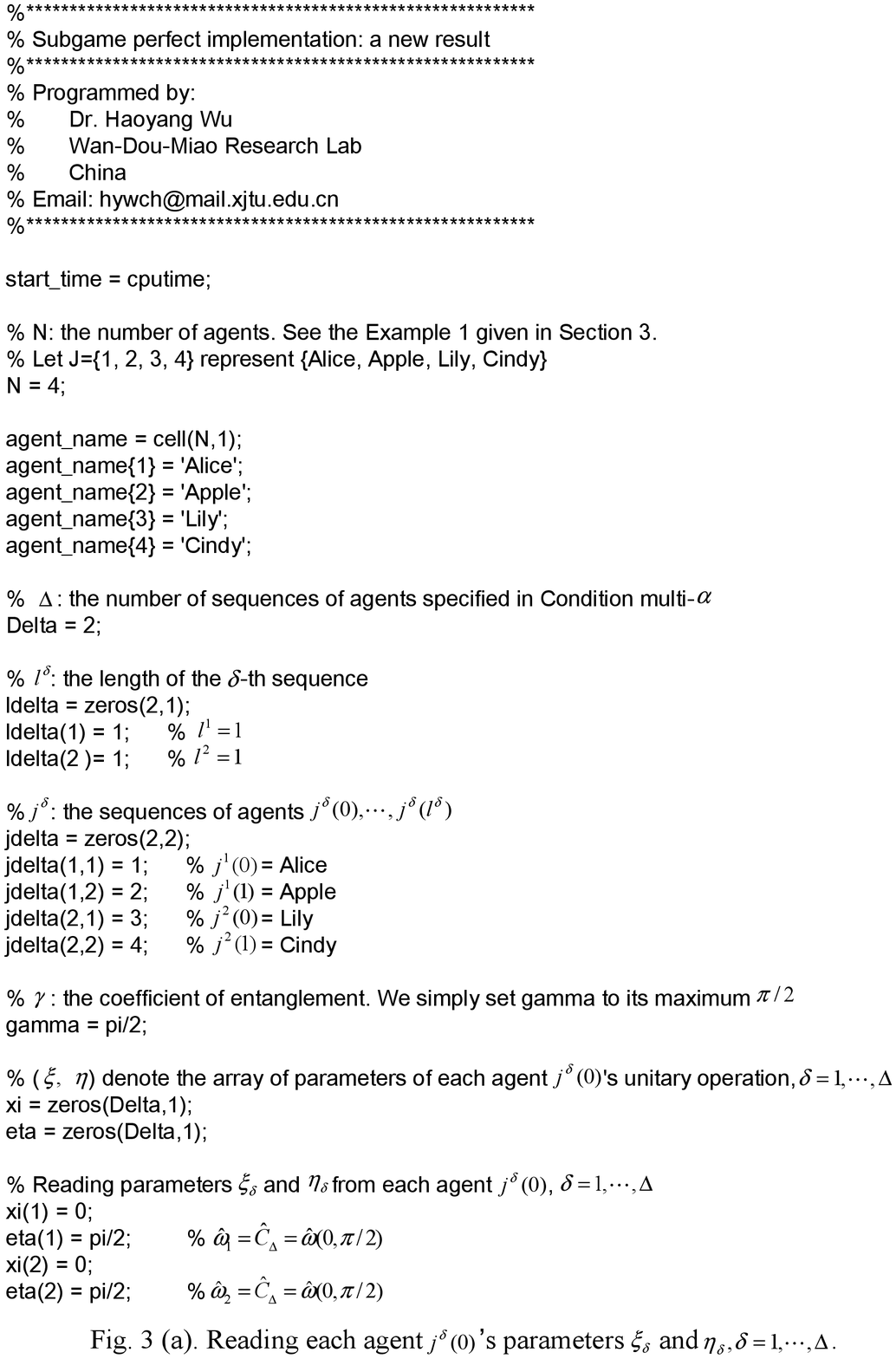}
\end{figure}

\begin{figure}[!t]
\centering
\includegraphics[height=5.2in,clip,keepaspectratio]{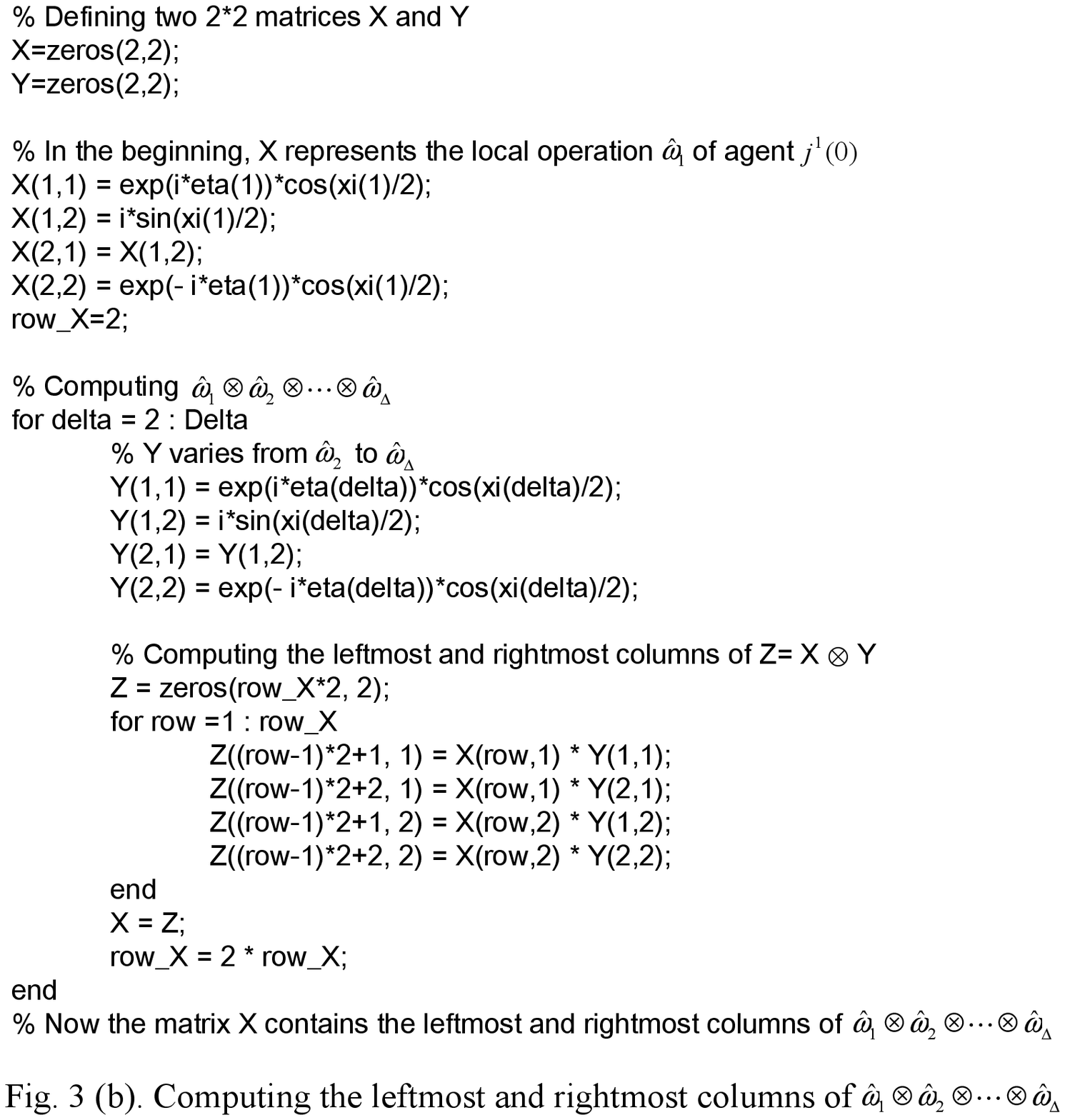}
\end{figure}

\begin{figure}[!t]
\centering
\includegraphics[height=2.9in,clip,keepaspectratio]{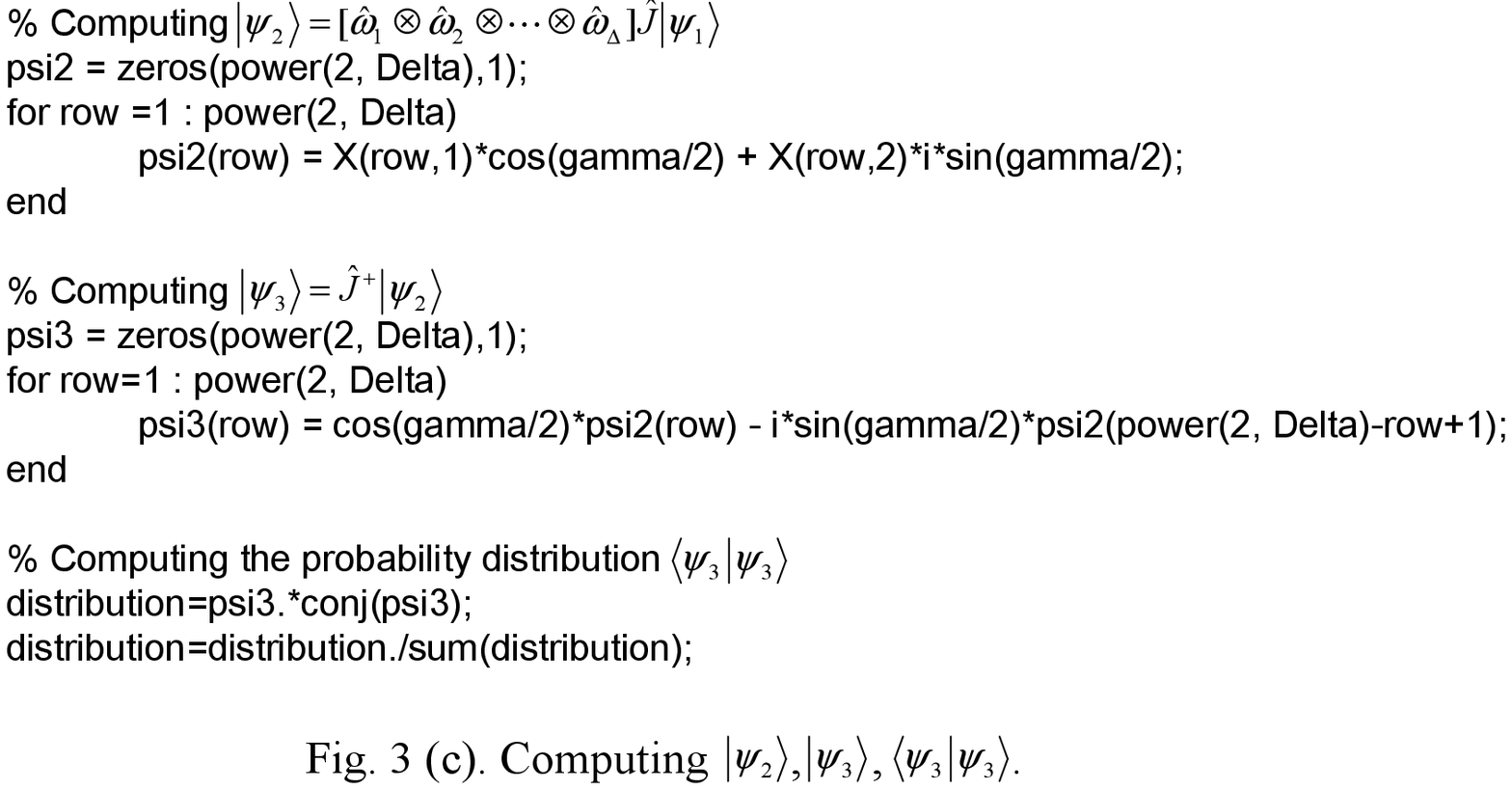}
\end{figure}

\begin{figure}[!t]
\centering
\includegraphics[height=8.5in,clip,keepaspectratio]{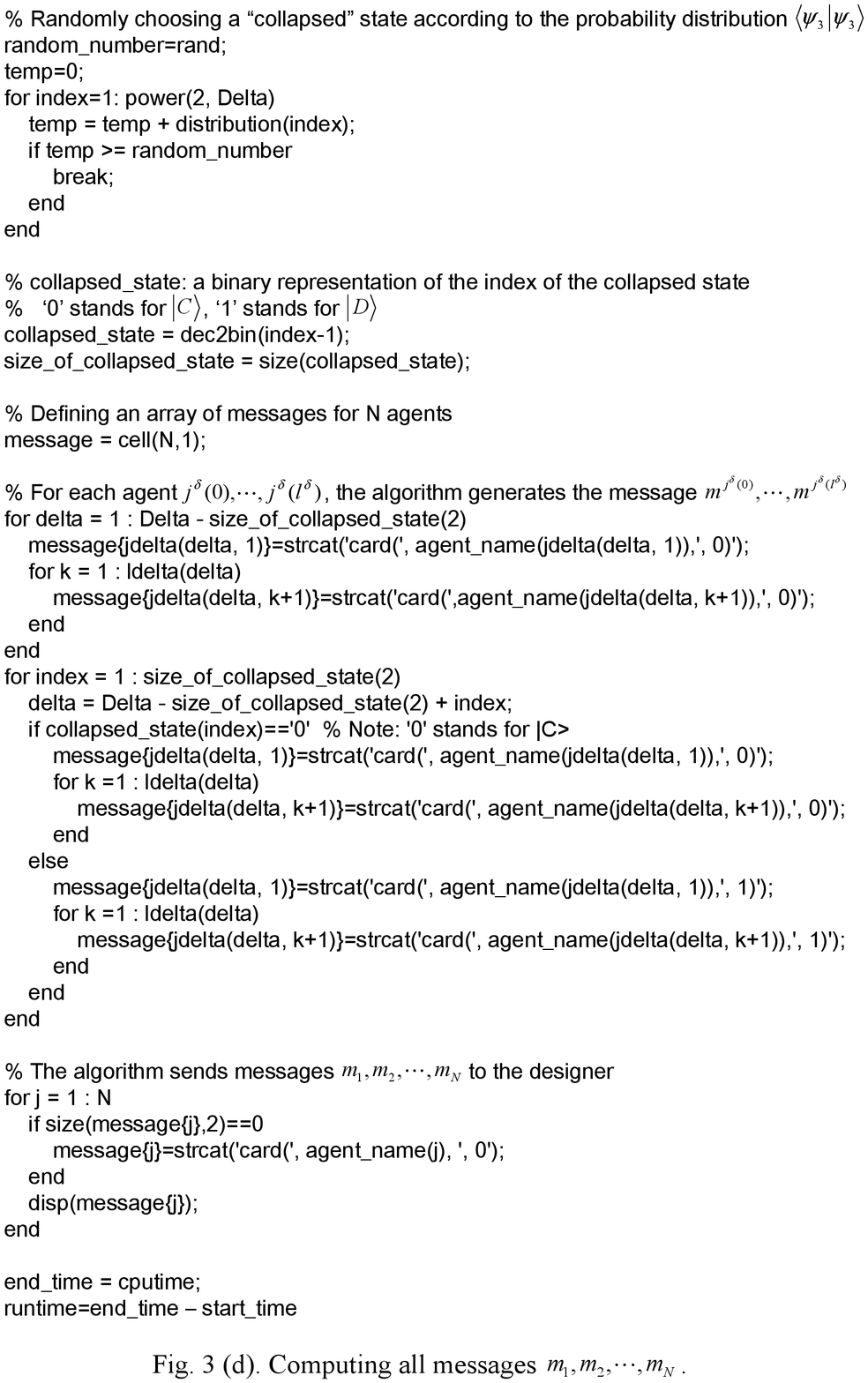}
\end{figure}
\end{document}